\documentclass[a4paper]{article}
\usepackage{amscd,amsmath,amssymb}
\usepackage[usenames, dvipsnames]{color}

\newcommand{\dimg}{{\mbox dim }\mathfrak{g}}

\newcommand{\p}[1]{(\ref{#1})}

\newcommand{\halpha}{{\hat \alpha}}
\newcommand{\hbeta}{{\hat \beta}}
\newcommand{\hgamma}{{\hat \gamma}}

\newcommand{\Tr}{\textrm{Tr}}
\newcommand{\ad}{{\sf ad}}

\newcommand{\be}{\begin{equation}}
\newcommand{\ee}{\end{equation}}
\newcommand{\bea}{\begin{eqnarray}}
\newcommand{\eea}{\end{eqnarray}}

\newcommand{\ba}{\begin{array}} \newcommand{\ea}{\end{array}}

\newcommand{\nn}{\nonumber}
\def\theequation{\arabic{section}.\arabic{equation}}

\begin{document}
\begin{flushright}
%	\today\\
%	Version 1.0
\end{flushright}\vspace{1cm}
\begin{center}
	{\Large\bf  The uniform structure of $\mathfrak{g}^{\otimes 4}$ }
\end{center}
\vspace{1cm}

\begin{center}
{\Large\bf M.~Avetisyan$^1$, A.P.~Isaev$^2$, S.O.~Krivonos$^2$,
 R.~Mkrtchyan$^1$}
\end{center}

\vspace{0.2cm}

\begin{center}
	\vspace{0.3cm}
		{\it 1. Alikhanian National Science Laboratory (Yerevan Physics Institute),\\ Alikhanian Br. 2, Yerevan 0036, Armenia}
	\vspace{0.3cm}
	{\it
	2. Bogoliubov  Laboratory of Theoretical Physics, JINR,
	141980 Dubna, Russia
}
	\vspace{0.5cm}
	
{\tt  maneh.avetisyan@gmail.com,isaevap@theor.jinr.ru, krivonos@theor.jinr.ru,  mrl55@list.ru, }
\end{center}
\vspace{2cm}

\begin{abstract}
	\noindent  We obtain a uniform decomposition into  Casimir eigenspaces (most of which are irreducible) of the fourth power of the adjoint representation $\mathfrak{g}^{\otimes 4}$ for all simple Lie algebras.  We present  universal, in Vogel's sense, formulae for the dimensions and split Casimir operator's eigenvalues of all terms in this decomposition. We assume that  a similar uniform decomposition into Casimir eigenspaces with universal dimension formulae exists for an arbitrary power of the adjoint representations.
\end{abstract}

\newpage

\pagenumbering{arabic}
\setcounter{page}{1}

\section{Introduction}
It was observed by P.~Vogel \cite{Vogel} that symmetric $Sym^2\mathfrak{g}$ and anti-symmetric $\wedge^2\mathfrak{g}$ modules of the square of the adjoint representation
$\mathfrak{g}^{\otimes 2 }$ can be decomposed into irreducible representations in a uniform way for all simple Lie algebras. This observation was made in the more general framework of Vogel's Universal Lie Algebra but is particularly applicable to usual simple Lie algebras. More exactly,
\be\label{ad2}
Sym^2 \mathfrak{g}= 1+ Y_2+Y_2'+Y_2'', \qquad
 \wedge^2 \mathfrak{g} = \mathfrak{g} +X_2 \,,
\ee
where $Y_2,Y_2',Y_2'', X_2$ are the irreducible representations of a simple Lie group extended by the automorphism group of Dynkin diagram of the corresponding simple Lie algebra \cite{Deligne,Cohen}. In what follows, irreducibility will always be considered w.r.t. that extended group $G$.

It was shown \cite{Vogel} that the dimensions of  all constituents are the rational functions of three (universal, Vogel's) parameters  $\alpha, \beta, \gamma$:

\bea\label{repsad2}
dim \, \mathfrak{g} &=&  \frac{(\halpha-1)(\hbeta-1)(\hgamma-1)}{\halpha \hbeta\hgamma} \\
dimX_2 &=& \frac{1}{2} \dimg(\dimg-3)= \frac{(2\halpha+1)(2\hbeta+1)(2\hgamma+1)(1-\halpha)(1-\hbeta)(1-\hgamma)}{8 \halpha^2 \hbeta^2 \hgamma^2}\\
 dimY_2&=&-\frac{(3\halpha-1)(\hbeta-1)(\halpha+\hbeta-1)(\hgamma-1)(\halpha+\hgamma-1)}{2 \halpha^2 (\halpha-1)\hbeta(\halpha-\hgamma)\hgamma}, \; \\
 dimY_2'&=& \left(dimY_2\right)_{\halpha \leftrightarrow \hbeta}, \,\,\, dimY_2''= \left(dimY_2\right)_{\halpha \leftrightarrow \hgamma}.\nn
\eea
Here,
\be
\halpha=\frac{\alpha}{2 t},\;\hbeta=\frac{\beta}{2 t},\;\hgamma=\frac{\gamma}{2 t},\quad t= \alpha+\beta+\gamma,\; \halpha+\hbeta+\hgamma=\frac{1}{2}
\ee

The meaning of these parameters is as follows: if we normalize the second Casimir operator to have eigenvalue $2t$ on the adjoint representation, with an arbitrary $t$, then its eigenvalues on the representations $Y_2,Y_2',Y_2''$ are $4t-2\alpha, 4t-2\beta, 4t-2\gamma$, respectively, and $\alpha+\beta+\gamma=t$. In the case when some representations have zero dimensions, one still considers them
 in the expansion (\ref{ad2}).
 %having these (non-zero) values.
 Taking into account the arbitrariness of the scale of the Casimir operator (arbitrary $t$), we see from definition  that Vogel's parameters are the homogeneous coordinates of the projective plane. The values of these parameters for simple Lie algebras are given in Vogel's Table 1.  In  this table, we  use the  normalization $t=1/2$ of the  Vogel parameters, different from those in \cite{Vogel, Landsberg}.

 \begin{center}\label{tab1}
	Table 1. \\ [0.2cm]
	\begin{tabular}{|c|c|c|c|c|c|c|c|c|}
		\hline
		$\;\;$ & $sl(N)$ & $so(N)$&  $sp(N)$ &
        $\mathfrak{g}_2$ & $\mathfrak{f}_4$
		& $\mathfrak{e}_6$ &
		$\mathfrak{e}_7$ & $\mathfrak{e}_8$  \\
		\hline
		$\halpha=\frac{\alpha}{2 t}$&\footnotesize $-1/N$ &\footnotesize $ -1/(N-2)$   &\footnotesize  $1/(N+2)$ &
		\footnotesize $-1/4$ &\footnotesize  $-1/9$ &\footnotesize  $-1/12$
		&\footnotesize  $-1/18$ &\footnotesize  $-1/30$  \\
		\hline
		$\hbeta=\frac{\beta}{2 t}$  &\footnotesize $1/N$ &
        \footnotesize $2/(N-2)$&\footnotesize $-2/(N+2)$ &
		\footnotesize $5/12$ &\footnotesize $5/18$ &\footnotesize $1/4$ &
		\footnotesize $2/9$ &\footnotesize $1/5$ \\
		\hline
		$\hgamma=\frac{\gamma}{2 t}$  &\footnotesize 1/2& \footnotesize $(N-4)/(2N-4)$ &\footnotesize $(N+4)/(2N+4)$&
		\footnotesize $1/3$ &\footnotesize $1/3$ &\footnotesize $1/3$ &
		\footnotesize $1/3$ &\footnotesize $1/3$ \\
		\hline
	\end{tabular}
\end{center}
In the same paper \cite{Vogel}, P.~Vogel found a uniform decomposition for the cube of the adjoint representation $\mathfrak{g}^{\otimes 3}$ into irreducible representations of the Universal Lie Algebra and calculated their dimensions in terms of the universal parameters $\halpha,\hbeta,\hgamma$; they
 recover the decomposition and dimensions of irreps of the corresponding simple Lie algebra. The dimensions are usually given by the formulae of type \p{repsad2}. However, the completely anti-symmetric part of this decomposition contains the representation $\mathbb{X}_3$
\be\label{ad3}
\wedge^3 \mathfrak{g} = 1 +X_2 +Y_2+Y_2'+Y_2''+ \mathbb{X}_3
\ee
which is Casimir eigenspace (i.e. its all vectors are eigenvectors of the second Casimir operator with the same eigenvalue) but not irreducible. Its dimension is easily found %from these formulae
to be
\be\label{dimx3}
dim\mathbb{X}_3 = \frac{1}{6} \dimg (\dimg-1)(\dimg-8)
\ee
which means that it is still universal, i.e. expressible  in terms of the Vogel parameters.
By introducing the fourth-order Casimir operator in the Universal Lie Algebra, Vogel decomposed this representation into three (at general values of the Vogel parameters) irreducible ones and calculated their dimensions. The corresponding formulae include three  additional parameters $\mu,\nu$ and $\lambda$. They actually are the functions of Vogel's parameters since they are the solutions of the third order equation over one unknown with coefficients being rational functions of the Vogel parameters $\alpha, \beta, \gamma$.

For the values of the parameters from the Vogel table,
the representation $\mathbb{X}_3$ is decomposed into one $X_3$ (for exceptional algebras, \cite{Vogel,Deligne,Cohen}) or two, for $sl$ and $so/sp$ algebras \cite{Vogel}, irreducible representations. More specifically, in the latter case, the representation
 $\mathbb{X}_3$ is the sum of two irreducible representations,  $\widetilde X_3$ and $\widehat X_3$, :
\bea
\mathbb{X}_3=\widetilde X_3+\widehat X_3
\eea
\bea
sl(N): & dim{\widehat X}_3 =\frac{1}{9}(N^2-1)^2(N^2-9), &
dim{\widetilde X}_3 =\frac{1}{18}(N^2-1)(N^2-4)(N^2-9), \\
so(N): & dim{\widehat X}_3 =\frac{1}{72}(N^2-16)(N-3)(N^2-1)N, &
dim{\widetilde X}_3 =\frac{1}{144}(N-5)(N^2-1)N^2(N+2).
\eea

In the paper \cite{Cohen}, the uniform structure of the decomposition of $\mathfrak{g}^{\otimes 4}$ into irreducible representation has been fully analyzed  for the exceptional algebras.
%, particularly it is decomposed into irreducible modules.
In this Note, we would like to find 
 analogous decomposition formulae for $\mathfrak{g}^{\otimes 4}$
 for all simple Lie algebras.
It is natural to restrict ourselves to decomposition into Casimir eigenspaces only, although in  most cases these subspaces are also irreducible
(w.r.t. the extended group $G$). With  this restriction, we avoid the need for complicated universal formulae for the dimensions of irreducible subspaces of representation $\mathbb{X}_3$, and we avoid the problem \cite{V} of the very existence of such formulae for other similar Casimir eigenspaces, such as $X_4$ below.

On the $n$-th power of the adjoint representation naturally acts the permutation group $S_n$ commuting with group $G$ action and correspondingly $\mathfrak{g}^{\otimes n}$ can be decomposed into the sum of irreps of these groups. Irreps of $S_n$  are given by partitions (or Young diagrams) $R$ with $n$ boxes, e.g., for $n=2$
 we have $R=(2,0), (1,1)$.   So the product
 $\mathfrak{g}^{\otimes 4}$ (which we are interested in this Note) can be expanded into five invariant subspaces, corresponding to the irreducible representations
 $R=(4), (3,1), (2,2), (2,1,1),(1,1,1,1)$ of $S_4$. Below we use the notation $[R]\mathfrak{g}$ proposed  in \cite{Deligne,Cohen}. So, these are: symmetric ${[(4)]\mathfrak{g}}$, anti-symmetric  ${[(1,1,1,1)]\mathfrak{g}}$, "window"  ${[(2,2)]\mathfrak{g}}$, hook  ${[(3,1)]\mathfrak{g}}$, and hook  ${[(2,1,1)]\mathfrak{g}}$
   invariant subspaces.

We  decompose these spaces into Casimir eigenspaces in the universal, a la Vogel \cite{Vogel}, form. Particularly, we present the universal dimension formulae for all the constituents of the decomposition. When universal parameters take
 the values from Vogel's table, these decompositions go into the decompositions of the corresponding simple Lie algebra, and dimension formulae give dimensions of the corresponding representations. For exceptional series our formulae coincide with those of \cite{Cohen}, providing an important check.

Below we present the final complete formulae for decompositions and universal dimensions. The more detailed description and derivations are postponed for extended paper.

\setcounter{equation}0
\section{Main tools: Split Casimir Operator}
One of the first questions in  constructing  decomposition formulae is
what calculations we have to perform to get such formulae. The method used by P.~Vogel in \cite{Vogel} in the framework of the Universal Lie Algebra is very complicated and also has some principal restrictions \cite{V}, so it is
rather difficult to extend it to higher powers of $\mathfrak{g}$. Classical direct calculations by weight space decomposition  appearing in decomposition formulae for  exceptional algebras \cite{Cohen}  have  a lot of simplifications in comparison with the general case with $sl, so, sp$ algebras leading to extremely short decomposition formulae.
Also, in the general case, we do not need such a detailed decomposition, since we need a decomposition into Casimir eigenspaces, only.

In a series of papers \cite{IsPr, IsKri1,IsKri2}, we advocated the use of the Split Casimir operator \cite{Book1} to construct  decomposition formulae. The (second) split Casimir operator can be defined for any simple complex Lie algebra $\mathfrak{g}$ with the basis elements $X_a$ as
\be\label{splitC}
{\widehat C_{(2)}} = g^{ab}\, X_a \otimes X_b,
\ee
where $g^{ab}$ is inverse of the Cartan-Killing metric $g_{ab}$ defined in a standard way as
\be\label{g}
g_{ab} \equiv  \Tr (\ad(X_a)\cdot \ad(X_b)).
\ee

Evidently, the split operator differs from the usual second Casimir operator acting on the tensor square of adjoint, by the constant multiplying by  unit operator. Similarly, one can introduce higher split ($n$-split) Casimir operators:

\be\label{hcas}
{\widehat C}_{(n)} = \sum_{i < j}^n {\widehat C}_{ij} ,
\ee
where
\be\label{hcas1}
{\widehat C}_{ij}=g^{ab} \left( I^{\otimes(i-1)}\otimes X_a \otimes
I^{\otimes(j-i-1)}\otimes X_b \otimes I^{\otimes(n-j)}\right) .
\ee

Note that our ${\widehat C}_{(2)}$ operator coincides, up to the sign, with Vogel's  \cite{Vogel} $\psi$ operator.
The main properties of the split Casimir operators, which help us to find  proper decomposition formulae, are:
\begin{itemize}
\item To find  decomposition formulae for $\mathfrak{g}^{\otimes n}$, we need
to use $n$-split Casimir operator (\ref{hcas});

\item The $n$-split Casimir operators obey the characteristic identity
\be
\prod_{j} \left( {\widehat C}_{(n)} +\lambda_j \right) =0 \; ,
\;\;\;\; (\lambda_i \neq \lambda_j \; \forall i \neq j) \; ,
\ee
where the product goes over %all
representations which appear in the expansion of $n$-power of the adjoint representation $\mathfrak{g}^{\otimes n}$
and we leave only the factors in which
values $\lambda_j$  are pairwise different. Here
$\lambda_j= \frac{1}{2} (c_{2}^{(\Lambda)} -  n)$ is the (minus of the) eigenvalue of the $n$-split Casimir operator in the subrepresentation $\Lambda$ in
$\mathfrak{g}^{\otimes n}$, $c_{2}^{(\Lambda)}$ is the value of the quadratic
Casimir operator in the representation $\Lambda$
(we use the normalization when $c_{2}^{(\mathfrak{g})}=1$).
For example, the projection of  ${\widehat C}^+_{(2)}$ on the symmetric subspace of the square of the adjoint obeys the equation
\be\label{exam1}
\left( {\widehat C}^+_{(2)}+1\right)\left( {\widehat C}^+_{(2)}+\halpha\right)\left( {\widehat C}^+_{(2)}+\hbeta\right)\left( {\widehat C}^+_{(2)} +\hgamma\right) =0.
\ee
Here each multiplier corresponds to the
subrepresentation in $[(2)]\mathfrak{g}$ (see \p{ad2}).
To calculate the dimensions of these representations, one has to know the
traces of higher powers of the corresponding projection of the split Casimir operator, in particular, for \p{exam1}, one has to know $\Tr\bigl(({\widehat C}^{+}_{(2)})^i\bigr), \; i=1,2,3$ (see \cite{IsKri1},\cite{IsKri2}).
\item We assume that the traces of higher powers of the
$R$-symmetrized projections ${\widehat C}_{(n)}^{R}$ of the $n$-split Casimir operator can
be expressed in terms of Vogel's parameters as
   $$
    {\rm Tr}\bigl(({\widehat C}_{(n)}^{R})^k\bigr) = \dim \mathfrak{g} \;
    \sum_{\ell =0}^{[\frac{k}{2}]-1}
    (\hat{\alpha}\hat{\beta}\hat{\gamma})^\ell \;
    P_\ell^{(R,k)}(\dim \mathfrak{g}) \; ,
    $$
    where $[p]$ is the integer part of $p$ and
    $P_\ell^{(R,k)}$ are polynomials in $\dim \mathfrak{g}$ of degree
   $(\ell+1)$.
    This will lead to universality of eigenvalues of the Casimir operator, since they are solutions to the characteristic equation. This hypothesis is partially confirmed below in the  decomposition formulae for $\mathfrak{g}^{\otimes 4}$ into Casimir eigenspaces, which we will discuss  in the next Section
   (see also \cite{IsKri2} for the case $\mathfrak{g}^{\otimes 3}$).
\end{itemize}

\setcounter{equation}0
\section{Decomposition formulae for $\mathfrak{g}^{\otimes 4}$}

\subsection{Anti-symmetric module $[(1,1,1,1)]\mathfrak{g}$}
The anti-symmetric part of $\mathfrak{g}^{\otimes 4}$ includes the following representations:

\be \label{as4}
[(1,1,1,1)]\mathfrak{g}=\mathfrak{g}\oplus X_2 \oplus \mathbb{X}_3 \oplus X_4\oplus C\oplus C'\oplus C''\oplus B\oplus B'\oplus B'' \oplus Y_2 \oplus Y_2' \oplus Y_2'' .
\ee
The dimensions of these representations read (the dimension of $X_2, Y_2,Y_2',Y_2'' $ and $\mathbb{X}_3$ are presented in \p{repsad2} and \p{dimx3})
\bea \label{as4dim}
 dimX_4 &=& \frac{1}{24} \dimg \left( \dimg-1\right)\left(\dimg-3\right)\left( \dimg-14 \right),  \nn \\
dim \mathbb{X}_3 & = &\frac{1}{6} \dimg\left( \dimg-1\right) \left(\dimg-8\right) , \nn \\
  dimB &=& -\frac{(\halpha-1)(\hbeta-1)(2\halpha+\hbeta)(2 \hbeta +1)(3\hbeta-1)(\hgamma-1)(2\halpha+\hgamma)(2\hgamma+1)(3\hgamma-1)}{8\, \halpha^2\,(\halpha-\hbeta)\, \hbeta^2\, (\hbeta-2\hgamma)(\halpha-\hgamma)(2\hbeta-\hgamma)\, \hgamma^2}, \nn \\
 dimB' &=& dim B_{\halpha \leftrightarrow \hbeta} , \,\,\,
 dimB'' = dim B_{\halpha \leftrightarrow \hgamma} \nn \\
 dimC & = & -\frac{2 (2\halpha+1)(\hbeta-1)(2\hbeta+1)(\hgamma-1)(\hbeta+\hgamma)(2 \hbeta+\hgamma)(2 \hgamma +1)(\hbeta+ 2\hgamma)}{3\, \halpha^3\,(\halpha-2 \hbeta)(\halpha-\hbeta)\, \hbeta\,(\alpha-2\hgamma)(\halpha-\hgamma)\, \hgamma}, \nn \\
 dimC' &=& dim C_{\halpha \leftrightarrow \hbeta} , \,\,\,
dimC'' = dim C_{\halpha \leftrightarrow \hgamma}
 \eea

Let us note that the parameters $\halpha, \hbeta, \hgamma$ are on an equal footing, so the whole theory is invariant (covariant) w.r.t. their permutations group $S_3$ \cite{Vogel}.  If we have some decomposition with universal dimension formulae, say \p{ad2}, and we have universal dimension formula for, e.g., $Y_2$, then formulae for the dimensions of $Y_2', Y_2''$ can be obtained from the permutation of the parameters
$\halpha, \hbeta, \hgamma$. In this way, we can obtain either additional two formulae, if the initial one was symmetric w.r.t. the switch of two parameters, or five additional formula, if the initial one was not symmetric under the switch of any two parameters. We will often characterize these representations in the following way: one of these representations, say $Y_2$, is Cartan square of adjoint (i.e. irrep with the highest weight equal to twice that of adjoint) which evidently should be in decomposition \p{ad2}. Then we shall say that other two appear from this one under permutation of parameters, as 
 this shows up in  their dimension formulae.

Almost all representations in decomposition \p{as4} appeared already in \cite{Vogel}, the new one is $X_4$. It is the representation from the series of representations $\mathbb{X}_k$ , described in (\cite{Deligne}, \cite{westb1} and references therein), which gives representations $X_2, \mathbb{X}_3$ at $k=2,3$. At an arbitrary $k$, it is Casimir space but generally not irreducible at $k \geq 2$ for $sl/sp/so$ algebras. For  exceptional algebras, they are irreducible.

Different representations in decomposition \p{as4} can be a true representation, zero, or virtual one, i.e. a representation with negative dimension,
actually cancelling the same true representation in other terms of this decomposition. All these possibilities are reflected in the dimension formulae \p{as4dim} when giving positive, zero, or negative values.

As we see from \p{as4dim}, all representations in the decomposition \p{as4} have universal formulae for their dimensions, so this is a desired generalization of Vogel's universal decompositions for $\mathfrak{g}^{\otimes 4}$, 
in the antisymmetric case. The only important difference is that our decomposition is not into irreducible
 subspaces but into Casimir eigenspaces, although, of course, most of these Casimir eigenspaces are actually irreducible ones, w.r.t. the extended group $G$.

Different representations in \p{as4} can be characterized as follows: $B, B', B''$ are the Cartan product of adjoint with $Y_2$, and its permutations; $C, C',C''$ are the Cartan product of adjoint with $X_2$ and its permutations.

The sum of the  dimensions of all these representations can be checked, at  arbitrary values of parameters, to be
\be
dim\,[(1,1,1,1)]\mathfrak{g} \, = \,\frac{1}{24} \dimg\,\left(\dimg-1\right)\left(\dimg-2\right)\left(\dimg-3\right),
\ee
as it should be.

In the case of exceptional algebras, we have
\be\label{ex1}
dimB=dimB'=dimY_2''=0, \quad dimC''= - dimX_2.
\ee
Therefore, with these cancellations,  we have for the exceptional algebras:
\be
[(1,1,1,1)]\mathfrak{g} \,=\mathfrak{g} \oplus  \mathbb{X}_3 \oplus X_4\oplus C\oplus C'\oplus B'' \oplus Y_2 \oplus Y_2' ,
\ee
which coincides with the result of  ref. \cite{Cohen} after identification of $A$ in \cite{Cohen} with $B''$.

\subsection{Symmetric module $[(4)]\mathfrak{g}$}
The symmetric part of $\mathfrak{g}^{\otimes 4}$ includes the following representations:
\bea
[(4)]\mathfrak{g} &= & 2 \oplus J\oplus J' \oplus J'' \oplus X_2 \oplus {\mathbb Z}_3  \oplus 3 Y_2 \oplus 3 Y_2' \oplus 3 Y_2''\oplus C\oplus C'\oplus C''\oplus Y_4\oplus Y_4'\oplus Y_4'' \oplus \nn \\
&& D \oplus D' \oplus D'' \oplus D''' \oplus D'''' \oplus D'''''.
\eea

Here
\bea
\mathbb{Z}_3&=& 2 \widehat X_3  \quad\mbox{for  } sl(N), \nn \\
\mathbb{Z}_3&=& \mathbb{X}_3  \quad\mbox{for  } so(N)  \mbox{ and exceptional algebras}
\eea	

The representation $J$ and its primes are Cartan square of $Y_2$  and its permutations. The representation $D$ and its primes are Cartan product of two adjoints and $Y_2$ and its permutations in $\halpha,\hbeta,\hgamma$. The representations $Y_4$ and primes are the fourth Cartan power of adjoint and its permutations
in $\halpha,\hbeta,\hgamma$. The universal dimension formulae of these representations are given in \cite{Landsberg,M16QD} and presented below:

\bea
 dimJ & = &
\frac{(\halpha +\hbeta ) (\halpha +\hgamma ) (2 \halpha
   +\hbeta -\hgamma ) (2 \halpha +2 \hbeta -\hgamma ) (2
   \halpha -\hbeta +\hgamma )
   (\halpha +2 \hbeta +\hgamma ) }{4\, \halpha ^2 \hbeta ^2
   \hgamma ^2 (\halpha -\hbeta ) (\halpha -\hgamma )
   (\hbeta -2 \hgamma ) (\hbeta -\hgamma )^2 (2 \hbeta
   -\hgamma ) (\halpha -\hbeta -\hgamma )}\times \nn \\
   && (2 \halpha +2 \hbeta
   +\hgamma ) (2 \halpha -\hbeta +2 \hgamma ) (\halpha
   +\hbeta +2 \hgamma ) (2 \halpha +\hbeta +2 \hgamma )
   (\halpha +2 \hbeta +2 \hgamma ), \nn \\
dimJ' & = & dimJ_{\halpha \leftrightarrow \hbeta} , \,\,\,
 dimJ''  =  dimJ_{\halpha \leftrightarrow \hgamma} , \nn \\
  dim\,\mathbb{Z}_3 & = & 2 dim{\widehat X}_3 =\frac{2}{9} (N^2-1)^2(N^2-9) \quad\mbox{for  } sl(N) ,\nn \\
	 &=& dim\,\mathbb{X}_3=\frac{1}{6} \dimg(\dimg-1)(\dimg-8), \quad\mbox{for  } so(N) \mbox{ and exceptional algebras}, \nn \\
dimY_4 & = & -\frac{(\halpha-1)(2\halpha-1)(7\halpha-1)(\hbeta-1)(\halpha+\hbeta-1)(2\halpha+\hbeta-1)(3\halpha+\hbeta-1)(\hgamma-1)}{24 \halpha^4 (\halpha-\hbeta)(2\halpha-\hbeta)(3\halpha-\hbeta)\hbeta(\halpha-\hgamma)(2\halpha-\hgamma)(3\halpha-\hgamma)\hgamma} \times  \nn \\
&& (\halpha+\hgamma-1)(2\halpha+\hgamma-1)(3\halpha+\hgamma-1), \nn\\
 dimY_4' & = &  \left(dimY_4\right)_{\halpha \leftrightarrow\hbeta}, \,\,\,
 dimY_4''  =   \left(dimY_4\right)_{\halpha \leftrightarrow\hgamma}, \nn \\
 dimD & = & \frac{(3\halpha-2\hbeta-2\hgamma)(\halpha-\hbeta-2\hgamma)(\hbeta+\hgamma)(\halpha+\hbeta+\hgamma)(2\halpha+\hbeta+\hgamma)(2\hbeta+\hgamma)(\halpha+2\hbeta+\hgamma)}{
\halpha^3(\halpha-\hbeta)^2(3\halpha-\hbeta)\hbeta^2(\halpha-2\hgamma)(\halpha-\hgamma)(2\halpha-\hgamma)(\hbeta-\hgamma)\hgamma} \times \nn \\
&& (2\halpha+2\hbeta+\hgamma)(\halpha+2\hgamma)(2\halpha-\hbeta+2\hgamma)(\halpha+\hbeta+2\hgamma)(2\halpha+\hbeta+2\hgamma)(\halpha+2\hbeta+2\hgamma), \\
dimD'& =& (dimD)_{\halpha\leftrightarrow \hbeta}, \; dimD''= (dimD)_{\halpha\leftrightarrow \hgamma}, \; dimD'''=(dimD)_{\hbeta\leftrightarrow \hgamma}, \nn \\
dimD'''' & = & (dimD)_{\halpha \rightarrow\hbeta\rightarrow \hgamma\rightarrow\halpha},  \;
 dimD''''' = (dimD)_{\halpha \rightarrow\hgamma\rightarrow \hbeta\rightarrow\halpha}  .
\eea
The dimension of $\mathbb{Z}_3$ can also be represented in the universal form:

\be\label{uf1}
dim\mathbb{Z}_3 = dim\mathbb{X}_3+\frac{1}{2}(\dimg+3) \frac{dimB\, dimB'\, dimB''}{dimY_2\, dimY_2'\, dimY_2''} +diff,
\ee
  where
\bea\label{diff}
diff& = &-\frac{16 (\halpha +\hbeta ) (2 \halpha +\hbeta ) (\halpha
   +2 \hbeta ) (\halpha +\hgamma ) (2 \halpha +\hgamma )
   (\halpha +2 \hgamma ) (\hbeta +\hgamma ) (2 \hbeta
   +\hgamma ) (\hbeta +2 \hgamma )} {\halpha ^2 \hbeta ^2 \hgamma ^2 (\halpha -2 \hbeta ) (2 \halpha -\hbeta ) (\halpha -2 \hgamma ) (2
   \halpha -\hgamma ) (\hbeta -2 \hgamma ) (2 \hbeta
   -\hgamma ) (\halpha -\hbeta -\hgamma ) } \times \nn \\
&& \frac{(\halpha -2 \hbeta -2
   \hgamma ) (2 \halpha +2 \hbeta -\hgamma ) (2 \halpha +2
   \hbeta +\hgamma ) (2 \halpha -\hbeta +2 \hgamma ) (2
   \halpha +\hbeta +2 \hgamma ) (\halpha +2 \hbeta +2
   \hgamma )}{(\halpha +\hbeta -\hgamma ) (\halpha -\hbeta +\hgamma )} .
\eea
Evidently,  $diff=0$ for all simple algebras.

As a check, the sum of dimensions of  all representations is
\be
dim{[4]\mathfrak{g}} \, = \,\frac{1}{24} \dimg\,\left(\dimg+1\right)\left(\dimg+2\right)\left(\dimg+3\right).
\ee

In the case of exceptional algebras, besides relations \p{ex1}, we have a lot of representations with
zero and negative dimensions:
\be\label{ex2}
dimJ=dimJ'=dimD'''=dimD''''=0,
\ee
\be\label{ex3}
dimY_4''=-1=-dimX_0, \; dimD'''''=-dimY_2, \;
dimD''=-dimY_2' .
\ee

Therefore, the decomposition for the exceptional algebras acquires the form
\be
[(4)]\mathfrak{g} =  1   \oplus 2 Y_2 \oplus 2 Y_2' \oplus C\oplus C' \oplus {\mathbb X}_3 \oplus D \oplus D'\oplus J''\oplus Y_4\oplus Y_4'
\ee
in full agreement with the result from \cite{Cohen} after identification
$J'' = J$.

\subsection{"Window" module $[(2,2)]\mathfrak{g}$}
The decomposition of the $\mathfrak{g}^{\otimes 4}$ for the "window" module $[(2,2)]\mathfrak{g}$ has the form:
\bea\label{winso}
[(2,2)]\mathfrak{g}= 2  \oplus E \oplus E'\oplus E''\oplus J \oplus J' \oplus J''\oplus \mathbb{X}_3\oplus\mathbb{Z}_3 \oplus 2 X_2 \oplus H \oplus H' \oplus H ''\oplus  \nn \\ 4 Y_2 \oplus 4 Y_2' \oplus 4 Y_2''  \oplus
 B \oplus B' \oplus B'' \oplus 2 C \oplus 2 C' \oplus 2 C'' \oplus D \oplus D' \oplus D''\oplus D''' \oplus D'''' \oplus D''''' .
\eea

The representations $E,E',E''$ are permutations of the Cartan product of the adjoint and representations $C$.
 The universal dimensions for the representations $E,E',E''$ are new and given by:
\bea
dimE& = & \frac{64 (\halpha +\hbeta ) (2 \halpha +\hbeta ) (\halpha +2 \hbeta ) (\halpha +\hgamma ) (2 \halpha +\hgamma ) (\halpha +2 \hgamma ) (\halpha
	+\hbeta +\hgamma ) (\halpha +2 \hbeta +\hgamma ) }{\halpha  \hbeta ^2 \hgamma ^2 (\halpha -\hbeta ) (\halpha -\hgamma ) (\hbeta -3 \hgamma ) (\hbeta -\hgamma )^2 (3 \hbeta -\hgamma ) (2 \halpha -\hbeta -\hgamma )} \times \nn \\
&&  (2 \halpha +2 \hbeta +\hgamma )(\halpha +\hbeta +2 \hgamma ) (2 \halpha +\hbeta +2 \hgamma )(\halpha +2 \hbeta +2 \hgamma ), \nn \\
dimE' & = & dimE_{\halpha \leftrightarrow \hbeta},  \quad dimE''  =  dimE_{\halpha \leftrightarrow \hgamma}.
\eea

The representation $\mathbb{Z}_3$ is the Casimir one and reads:
\bea
\mathbb{Z}_3  &= & 3\, {\widehat X}_3 + \widetilde{X}_3 \quad
\mbox{ for } sl(N), \nn\\
\mathbb{Z}_3 & = & 2\, \mathbb{X}_3  \quad
\mbox{ for } so(N) \mbox{  and exceptional algebras}.
\eea
As already presented for the symmetric case,  $dim\mathbb{Z}_3$ can be represented in the universal form:
\be
dim\mathbb{Z}_3= 2 dim\mathbb{X}_3+\frac{1}{2}(\dimg+3) \frac{dimB\, dimB'\, dimB''}{dimY_2\, dimY_2'\, dimY_2''} +diff.
\ee

Finally, the new representation $H$ and its primed versions, is the Cartan square of $X_2$ and its permutations. They have the following universal dimensions \cite{AM18}:

\bea
dimH & = & \frac{(\halpha+2\hbeta)(3\halpha-2\hbeta-2\hgamma)(\halpha-\hbeta-2\hgamma)(\halpha-2\hbeta-\hgamma)(2\halpha+\hbeta+\hgamma)(2\hbeta+\hgamma)}{12 \halpha^4 (\halpha-2\hbeta)(\halpha-\hbeta)^2\hbeta^2(\halpha-2\hgamma)(\halpha-\hgamma)^2(2\halpha-\hbeta-\hgamma)\hgamma^2} \times  \nn \\
&& (\halpha+2\hbeta+\hgamma)(2\halpha+2\hbeta+\hgamma)(\halpha+2\hgamma)(\hbeta+2\hgamma)
(\halpha+\hbeta+2\hgamma)(2\halpha+\hbeta+2\hgamma)(\halpha+2\hbeta+2\hgamma), \nn\\
 dimH' & = &  dim\left( H\right)_{\halpha \leftrightarrow\hbeta}, \,\,\,
 dimH''  =   dim\left( H\right)_{\halpha \leftrightarrow\hgamma}.
\eea

The sum of dimensions of  all representations is as expected:
\be
dim [(2,2)]\mathfrak{g} = \frac{1}{12} \dimg^2 (\dimg-1)(\dimg+1) .
\ee

In the case of exceptional algebras, besides the relations \p{ex1},\p{ex2}, \p{ex3},
we also have the following new ones:
\be\label{ex4}
dimH'' = dimX_2, \; dimE' =-dimC, \; dim E = - dimC'.
\ee
Therefore, the decomposition of the "window" part of $\mathfrak{g}^{\otimes 4}$
for the exceptional algebras reads

\bea\label{winex}
[(2,2)]\mathfrak{g}&=& 2  \oplus E'' \oplus J \oplus X_3 \oplus  X_2 \oplus H \oplus H' \oplus 2 Y_2 \oplus 2 Y_2'   \oplus B'' \oplus  C \oplus  C' \oplus  D \oplus D'
\eea
in  full agreement with the result presented in \cite{Cohen}.

\subsection{"Hook" modules}

There are two hook diagrams:
$[(3,1)]\mathfrak{g}={\tiny
\begin{tabular}{|c|c|c|}
\hline
$\!\!\!$ & $\!\!\!\!$ & $\!\!\!\!$\\
\hline
$\!\!\!$ &   \multicolumn{2}{c}{\!\!\!\!} \\
\cline{1-1}
%$\!\!\!$ &   \multicolumn{1}{c}{\!\!\!} \\
%\cline{1-1}
\end{tabular} }$ and
$[(2,1,1)]\mathfrak{g}= {\tiny
\begin{tabular}{|c|c|}
\hline
$\!\!\!$ & $\!\!\!$ \\
\hline
$\!\!\!$ &   \multicolumn{1}{c}{\!\!\!} \\
\cline{1-1}
$\!\!\!$ &   \multicolumn{1}{c}{\!\!\!} \\
 \cline{1-1}
\end{tabular} }$.

\subsubsection{Hook module $[(3,1)]\mathfrak{g}$}
The decomposition of the   hook $[(3,1)]\mathfrak{g}$ module has the form:
\bea\label{h1}
[(3,1)]\mathfrak{g}&=&  3\,\mathfrak{g} \oplus E \oplus E'\oplus E''\oplus\mathbb{K}_3 \oplus 6\, X_2 \oplus G \oplus G' \oplus G''
\oplus F \oplus F' \oplus F''\oplus F''' \oplus F'''' \oplus F'''''\oplus\nn \\
&& 3\, Y_2 \oplus 3\, Y_2' \oplus 3\, Y_2''  \oplus
 3\,B \oplus 3\, B' \oplus 3\, B'' \oplus 3 C \oplus 3 C' \oplus 3 C'' \oplus Y_3 \oplus Y_3' \oplus Y_3'' \oplus \nn\\
&& D \oplus D' \oplus D''\oplus D''' \oplus D'''' \oplus D'''''
\eea
Here the representation $\mathbb{K}_3$ is the Casimir one and reads:
\bea
\mathbb{K}_3  &= & 3\, {\widehat X}_3 + \widetilde{X}_3 \quad
\mbox{ for } sl(N), \nn\\
\mathbb{K}_3 & = & 2\, {\widehat X}_3 + \widetilde{X}_3 \quad
\mbox{ for } so(N), \nn\\
\mathbb{K}_3 & = & X_3  \quad
\mbox{ for exceptional algebras}.
\eea
Its dimension $dim\mathbb{K}_3$ has a very complicated  universal form, so we omit it.
However, it can easily be found as a consequence of relations  \p{h1} and \p{dim31} below, i.e. as a difference of $dim[(3,1)]\mathfrak{g}$ in \p{dim31} and the sum of universal dimensions in r.h.s. of \p{h1}
\be\label{dimK3}
dim\mathbb{K}_3 =dim[(3,1)]\mathfrak{g} -3\, dim\mathfrak{g} -dimE-dimE'-dimE''- 6\, dimX_2- \ldots .
\ee
The sum of dimensions of all representations is
\be\label{dim31}
dim[(3,1)]\mathfrak{g}= \frac{1}{8} \dimg (\dimg-1)(\dimg+1) (\dimg+2) ,
\ee
while the dimensions of the new irreducible representations $G$ and $F$ are presented below in
eqs. \p{dimG} and \p{dimF}.

The new representations $G,G',G''$ are the Cartan product of $X_2$ with two adjoints and permutations. Their  universal dimensions are \cite{AM19}:
\bea\label{dimG}
dimG & = &-\frac{(2 \hbeta +\hgamma ) (\hbeta +2 \hgamma ) (\halpha -2 \hbeta -2 \hgamma ) (\halpha -\hbeta -2 \hgamma ) (\halpha
   -2 \hbeta -\hgamma ) (\halpha +\hbeta +\hgamma ) (2 \halpha +\hbeta +\hgamma ) }{2 \halpha ^4 \hbeta  \hgamma  (\halpha -\hbeta )^2 (2 \halpha -\hbeta ) (\halpha -\hgamma )^2 (2 \halpha
   -\hgamma )}\times \nn \\
   && (\halpha +2 \hbeta +\hgamma ) (2
   \halpha +2 \hbeta +\hgamma ) (\halpha +\hbeta +2 \hgamma ) (2 \halpha +\hbeta +2 \hgamma ) (\halpha +2 \hbeta +2    \hgamma ),\nn \\
   dimG' & = & \left( dimG\right)_{\halpha \leftrightarrow \hbeta}, \quad
   dimG'' = \left( dimG\right)_{\halpha \leftrightarrow \hgamma}.
 \eea
Finally,  the representation $F$ and its primes are the Cartan product of $X_2$ and $Y_2$ and its permutations; for their dimensions we have the following (new) universal formulae:
\bea\label{dimF}
dimF & = & -\frac{(\halpha +\hgamma ) (2 \hbeta +\hgamma ) (\halpha -2 \hbeta -2 \hgamma ) (\halpha -\hbeta -2 \hgamma ) (\halpha +\hbeta +\hgamma )^2 (2 \halpha
   +\hbeta +\hgamma ) (\halpha +2 \hbeta +\hgamma ) }{\halpha ^3 \hbeta ^2 \hgamma ^2 (\halpha -\hbeta )^2 (2 \halpha -\hbeta ) (\halpha
   -\hgamma )^2 (\hbeta -\hgamma )}\times \nn \\
&& (2 \halpha +2 \hbeta +\hgamma ) (2 \halpha -\hbeta +2 \hgamma ) (\halpha +\hbeta +2 \hgamma ) (2 \halpha +\hbeta +2 \hgamma ) (\halpha +2 \hbeta +2 \hgamma ) , \nn \\
dimF'& =& (dimF)_{\halpha\leftrightarrow \hbeta}, \; dimF''= (dimF)_{\halpha\leftrightarrow \hgamma}, \; dimF'''=(dimF)_{\hbeta\leftrightarrow \hgamma}, \nn \\
dimF'''' & = & (dimF)_{\halpha \rightarrow\hbeta\rightarrow \hgamma\rightarrow\halpha},  \;
dimF''''' = (dimF)_{\halpha \rightarrow\hgamma\rightarrow \hbeta\rightarrow\halpha}  .
\eea

Passing to the exceptional algebras, we have to take into account, besides  relations \p{ex1}, \p{ex2}, \p{ex3}, \p{ex4}, multiple new relations between the representations:
\be\label{ex5}
\left\{dimG'', dimF'', dimF''', dimF'''',dimF''''\right\}=0, \quad
dim\mathfrak{g}+dimY_3''=0. \nn
\ee
Correspondingly, the decomposition formula is simplified to be
\bea\label{h1ex}
[(3,1)]\mathfrak{g}&=&  2\,\mathfrak{g} \oplus E'' \oplus \mathbb{X}_3 \oplus 3\, X_2 \oplus G \oplus G' \oplus F \oplus F' \oplus  2\, Y_2 \oplus 2\, Y_2' \oplus
  3\, B'' \oplus 2 C \oplus 2 C'  \oplus  \nn\\
&& Y_3 \oplus Y_3'  \oplus D \oplus D' ,
\eea
which coincides with  similar expression presented in \cite{Cohen}.

\subsubsection{Hook module $[(2,1,1)]\mathfrak{g}$}
The decomposition of the   hook $[(2,1,1)]\mathfrak{g}$ module has the form:
\bea\label{h2}
[(2,1,1)]\mathfrak{g}&=&  4\,\mathfrak{g} \oplus E \oplus E'\oplus E''\oplus \mathbb{L}_3 \oplus 7\, X_2 \oplus I \oplus I' \oplus I''
\oplus F \oplus F' \oplus F''\oplus F''' \oplus F'''' \oplus F'''''\oplus\nn \\
&&  Y_2 \oplus  Y_2' \oplus  Y_2''  \oplus
 4\,B \oplus 4\, B' \oplus 4\, B'' \oplus 3 C \oplus 3 C' \oplus 3 C'' \oplus Y_3 \oplus Y_3' \oplus Y_3'' .
\eea
Here the representation $\mathbb{L}_3$ (Casimir subspace) reads
\bea
\mathbb{L}_3  &= & 2\widetilde X_3+2\widehat X_3  \quad
\mbox{ for } sl(N), \nn\\
&& 2\, {\widehat X}_3 + \widetilde{X}_3 \quad
\mbox{ for } so(N), \nn\\
&& X_3  \quad
\mbox{ for exceptional algebras}.
\eea
Similarly to the previous cases, it has the universal dimension:
\be
 dim\mathbb{L}_3 =dim\mathbb{K}_3-\frac{1}{2}(\dimg+3) \frac{dimB\, dimB'\, dimB''}{dimY_2\, dimY_2'\, dimY_2''}-diff .
\ee
	
The new Casimir eigenspaces are now  $I, \, I'$ and  $I''$ with the  dimensions

\bea
dimI &= &-\frac{(\halpha +\hbeta +\hgamma ) (\halpha +2 \hbeta +\hgamma ) (2 \halpha +2 \hbeta +\hgamma ) (\halpha +\hbeta +2 \hgamma ) (2 \halpha +\hbeta +2   \hgamma ) }{2 \halpha ^4 \hbeta ^2 \hgamma ^2 (\halpha -3 \hbeta ) (\halpha -2 \hbeta ) (\halpha -\hbeta )
   (\halpha -3 \hgamma ) (\halpha -2 \hgamma ) (\halpha -\hgamma ) (\halpha -\hbeta -\hgamma )} \times \nn \\
&&\left(4 \halpha^{10}-14 \halpha^9 \hbeta -14 \halpha^9 \hgamma -32 \halpha^8 \hbeta^2-8 \halpha^8 \hbeta  \hgamma -32 \halpha^8 \gamma ^2+116 \halpha^7 \hbeta^3+259 \halpha^7 \hbeta^2 \hgamma +259 \halpha^7 \hbeta  \hgamma^2+\right. \nn \\
&&   116 \halpha^7 \hgamma^3+116 \halpha^6 \hbeta^4+116 \halpha^6 \hbeta^3 \hgamma +196 \halpha^6 \hbeta^2 \hgamma^2+116 \halpha^6 \hbeta\hgamma^3+116 \halpha^6 \hgamma^4-310 \halpha^5 \hbeta^5-1963 \halpha^5 \hbeta^4 \hgamma - \nn \\
&&   4762 \halpha^5 \hbeta^3 \hgamma^2-4762 \halpha^5 \hbeta^2\hgamma^3-  1963 \halpha^5 \hbeta  \hgamma^4-310 \halpha^5 \hgamma^5-296 \halpha^4 \hbeta^6-2634 \halpha^4 \hbeta^5 \hgamma -9263 \halpha^4\hbeta^4
\hgamma^2- \nn \\
&&   13706 \halpha^4 \hbeta^3 \hgamma^3-9263 \halpha^4 \hbeta^2
\hgamma^4-2634 \halpha^4 \hbeta \hgamma^5-296 \halpha^4 \hgamma^6+256 \halpha^3 \hbeta^7+500 \halpha^3 \hbeta^6 \hgamma -2938 \halpha^3 \hbeta^5 \hgamma^2- \nn \\
&&   10206 \halpha^3 \hbeta^4\hgamma^3-10206 \halpha^3 \hbeta^3 \hgamma^4-2938 \halpha^3 \hbeta^2 \hgamma^5+500 \halpha^3 \hbeta
\hgamma^6+256 \halpha^3 \hgamma^7+352 \halpha^2 \hbeta^8+4248 \halpha^2 \hbeta^7 \hgamma +\nn \\
&&  20252 \halpha^2 \hbeta^6 \hgamma^2+48516 \halpha^2 \hbeta^5
\hgamma^3+64320 \halpha^2 \hbeta^4 \hgamma^4+48516 \halpha^2 \hbeta^3 \hgamma^5+20252 \halpha^2 \hbeta^2 \hgamma^6+4248 \halpha^2 \hbeta
   \hgamma^7+\nn \\
&&   352 \halpha^2 \hgamma^8+96 \halpha \hbeta^9+2976 \halpha \hbeta^8 \hgamma +20712 \halpha \hbeta^7 \hgamma^2+65496 \halpha
   \hbeta^6 \hgamma^3+112896 \halpha  \hbeta^5 \hgamma^4+112896
   \halpha  \hbeta^4 \hgamma^5+\nn\\
&&   65496 \halpha  \hbeta^3 \hgamma^6+20712
   \halpha  \hbeta^2 \hgamma^7+2976 \halpha  \hbeta  \hgamma^8+
   96 \halpha  \hgamma^9+576 \hbeta^9 \hgamma +5184 \hbeta^8 \hgamma^2+19728  \hbeta^7 \hgamma^3+\nn\\
&&\left.   41472 \hbeta^6 \hgamma^4+52704 \hbeta^5 \hgamma^5+41472
\hbeta^4 \hgamma^6+19728 \hbeta^3 \hgamma^7+5184 \hbeta^2 \hgamma^8+576 \hbeta  \hgamma^9 \right), \nn \\
dimI'& = & (dimI)_{\halpha\leftrightarrow\hbeta}, \quad
dimI'' =  (dimI)_{\halpha\leftrightarrow\hgamma}.
\eea
	
These spaces are given by the Cartan product of $\mathbb{X}_3$ (actually irreps inside it) with the adjoint and its permutations.

The sum of dimensions of all representations is as it should be, for an arbitrary values of the parameters:
	\be
dim[(2,1,1)]\mathfrak{g}= \frac{1}{8} (\dimg-2)(\dimg-1)\dimg (\dimg+1).
	\ee

Remembering the previous relations \p{ex1}, \p{ex2}, \p{ex3}, \p{ex4}, \p{ex5} between the
representations for exceptional groups together with the new one
\be\label{ex6}
dimI''+dim B'' =0 ,
\ee
one can easily obtain the decomposition for the exceptional algebras, which agrees with \cite{Cohen}:
\bea\label{h2ex}
[(2,1,1)]\mathfrak{g}&=&  3\,\mathfrak{g} \oplus E'' \oplus \mathbb{X}_3 \oplus 4\, X_2 \oplus I \oplus I'
\oplus F \oplus F' \oplus  Y_2 \oplus  Y_2'  \oplus\nn \\
&&
 3\, B'' \oplus 2 C \oplus 2 C' \oplus Y_3 \oplus Y_3' .
\eea

\section{Conclusion}

The Vogel parametrization of simple Lie algebras by the points on projective plane already appears to be very useful in many areas of the theory of Lie algebras and their applications.  In the present paper, we have extended the applicability of this parametrization into the decomposition of the fourth power of the adjoint representation. The novel feature is the restriction on Casimir eigenspaces, which allows us to get a completely universal, uniform decomposition for all simple Lie algebras. In the process, new universal dimension formulae were obtained, which are in full agreement with the already existing ones on the exceptional line. In the Appendix, Table 2, we presented the eigenvalues of the 4-split Casimir operator for  all representations that appear in the decomposition of $\mathfrak{g}^{\otimes 4}$.

We suggest the existence of such decomposition for arbitrary powers of the adjoint representation, i.e., that it can be decomposed into Casimir eigenspaces with universal dimensions formulae for each subspace.

\section*{Acknowledgments}
MA and RM are partially supported by the
Science Committee of the Ministry of Science
and Education of the Republic of Armenia under contract 21AG-1C060.
The work of API is supported by the RNF grant
23-11-00311.

\setcounter{equation}0
\section*{Appendix}
\def\theequation{A.\arabic{equation}}

The eigenvalues of the 4-split Casimir operator on the representations in the
decomposition of $\mathfrak{g}^{\otimes 4}$ read \vspace{0.1cm} \\

\begin{center}{Table 2}
	\end{center}

\be\label{tab2}
\begin{tabular}{|c|c|c|c|c|c|} \hline\hline
$1$ &$ \mathfrak{g} $ & $X_2 $& $\mathbb{X}_3,\mathbb{Z}_3,\mathbb{K}_3,\mathbb{L}_3$ & $X_4$
 	& \\ \hline
1 & $-\frac{3}{2} $& -1 & $-\frac{1}{2}$ & 0 & \\ \hline \hline
$Y_2$ & $Y_2'$ & $Y_2''$ & $Y_3$ & $Y_3'$ & $Y_3''$ \\ \hline
$-1-\halpha$& $-1-\hbeta $ & $-1-\hgamma$ &$-\frac{1}{2}-3\halpha$& $-\frac{1}{2}-3\hbeta $ & $-\frac{1}{2}-3\hgamma$\\ \hline\hline
$Y_4$ & $Y_4'$ & $Y_4''$ & $B$ & $B'$ & $B''$\\ \hline
$-6\halpha$& $-6\hbeta $ & $-6\hgamma$ & $-1+\halpha$& $-1+\hbeta $ & $-1+\hgamma$\\ \hline\hline
$C$ & $C'$ & $C''$ &  $J$ & $J'$ & $J''$ \\ \hline
$-\frac{1}{2}-\frac{3}{2}\halpha$& $-\frac{1}{2}-\frac{3}{2}\hbeta $ & $-\frac{1}{2}-\frac{3}{2}\hgamma$ &$-2(\hbeta+\hgamma)$ & $-2(\halpha+\hgamma)$ & $-2 (\halpha+\hbeta)$\\ \hline\hline
$D$ & $D'$ & $D''$ & $D'''$  & $D''''$ & $D'''''$\\ \hline
$-3\halpha-\hbeta$&$-3\hbeta-\halpha$ &$-3\hgamma-\hbeta$ &$-3\halpha-\hgamma$ & $-3\hbeta-\hgamma$& $-3\hgamma-\halpha$\\ \hline\hline
$H$ & $H'$ & $H''$ & $E$ & $E'$& $E''$\\ \hline
$-3\halpha$ & $-3\hbeta$ & $-3\hgamma$ &$-\frac{3}{2}(\hbeta+\hgamma)$ & $-\frac{3}{2}(\halpha+\hgamma)$ & $-\frac{3}{2} (\halpha+\hbeta)$ \\ \hline\hline
$G$ & $G'$ & $G''$ & $I$ & $I'$& $I''$\\ \hline
$-4\halpha$ & $-4\hbeta$ & $-4\hgamma$ &$-2\halpha$ & $-2\hbeta$ & $-2 \hgamma$ \\ \hline\hline
$F$ & $F'$ & $F''$ & $F'''$  & $F''''$ & $F'''''$\\ \hline
$-2\halpha-\hbeta$&$-2\hbeta-\halpha$ &$-2\hgamma-\hbeta$ &$-2\halpha-\hgamma$ & $-2\hbeta-\hgamma$& $-2\hgamma-\halpha$\\\hline
\end{tabular}\nn
\ee


\begin{thebibliography}{99}
\addtolength{\itemsep}{-1pt}
\bibitem{Vogel} P.~Vogel,  \textit{The Universal Lie algebra}, preprint (1999), https://webusers.imj-prg.fr/\~{}pierre.vogel/grenoble-99b.pdf
\bibitem{Deligne} P.~Deligne, \textit{ La s\'{e}rie exceptionnelle des groupes de Lie}, C.R.~Acad.~Sci. 322 (1996) 321.
\bibitem{Cohen} A.M.~Cohen, R.~de~Man, \textit{Computational evidence for Deligne's conjecture regarding exceptional Lie groups}, C.R.~Acad.~Sci.~Paris 322 (1996) p.427.
\bibitem{Landsberg} J.M.~Landsberg, L.~Manivel, \textit{A universal dimension formula for complex simple
Lie algebras}, Adv. Math. 201 (2006) 379.
\bibitem{V}
P.Vogel,   \textit{Algebraic structures on modules of diagrams}, preprint (1995),  www.math.jussieu.fr/\~{}vogel/diagrams.pdf, J. Pure Appl. Algebra {\bf 215} (2011), no. 6, 1292-1339.
\bibitem{IsPr} A.P.~Isaev, and A.A.~Provorov,
 \textit{Projectors on invariant subspaces of representations
   $\ad^{\otimes 2}$ of Lie algebras $so(N)$ and $sp(2r)$ and Vogel parametrization},  TMF 206:1 (2021) 3-22, arXiv:2012.00746[math-ph].
\bibitem{IsKri1} A.P.~Isaev, and S.O.~Krivonos,
 \textit{Split Casimir operator for simple Lie algebras, solutions of Yang-Baxter equations and Vogel parameters},  J.~Math.~Phys. 62, 083503 (2021), arXiv:2102.08258[math-ph].
 \bibitem{IsKri2} A.P.~Isaev, S.O.~Krivonos, A.A.~Provorov,
"Split Casimir operator for simple Lie algebras in the cube of ad-representation and Vogel parameters",
Int.~J.~Mod.~Phys. A 38 (2023) 06n07, 235003, arXiv:2212.14761[math-ph]
 \bibitem{Book1} A.P.~Isaev, V.A.~Rubakov,
\textit{Theory of groups and symmetries: Finite Groups, Lie Groups, And Lie Algebras}, World Scientific (2018).
\bibitem{westb1}
Bruce~W.~Westbury, \textit{Universal characters from the MacDonald identities}, Advances in Mathematics, 202 (2006), 50 – 63
\bibitem{M16QD}
R.L.~Mkrtchyan, \textit{On Universal Quantum Dimensions}, arxiv:1610.09910, Nuclear Physics B921,  2017, pp. 236-249,
\bibitem{AM18}	M.Y.~Avetisyan, R.L.~Mkrtchyan,  \textit{$X_2$ series of universal quantum dimensions}, J.~Phys.~A: Math.~Theor. (2020), Volume 53, Number 4, 045202;, arXiv:1812.07914[hep-th].
\bibitem{AM19}	M.Y.~Avetisyan, R.L.~Mkrtchyan,  \textit{On $(ad)^n(X_2)^k$ series of universal quantum dimensions},  J.~Math.~Phys. 61, 101701 (2020); arXiv:1909.02076[math-phys].


\end{thebibliography}
\end{document}